\documentclass{article}
\usepackage[utf8]{inputenc}
\usepackage[normalem]{ulem}
\usepackage{graphicx} 
\usepackage{xcolor}
\usepackage{tablefootnote}
\usepackage{cite}
\usepackage{amsmath,amssymb,amsfonts}
\usepackage{mathtools}
\usepackage{algorithmic}
\usepackage{textcomp}
\usepackage{relsize}

\usepackage[font=small,labelfont=bf]{caption}
\usepackage[square,numbers]{natbib}
\usepackage{hyperref}

\title{Efficient Dynamic Mesh Refinement Technique for Simulation of HPM Breakdown induced Plasma Pattern Formation}
\author{Pratik Ghosh and Bhaskar Chaudhury \thanks{Corresponding Author :pratik\_ghosh@daiict.ac.in; bhaskar\_chaudhury@daiict.ac.in}\\
\emph{Group in Computational Science and HPC} \\ \emph{DA-IICT, Gandhinagar - 382007 , India.}}
\date{ }
\providecommand{\keywords}[1]{\textbf{\textit{Keywords-}} #1}
\begin{document}
\maketitle
\begin{abstract}
Numerical simulation of the complex plasma dynamics associated with high power, high frequency microwave breakdown at high pressures, leading to the formation of filamentary plasma structures such as self-organized plasma arrays, is a computationally challenging problem. 
The widely used 2D EM-plasma fluid model, which accurately captures the experimental observations, requires a run-time of several days to months to simulate standard problems due to stringent numerical requirements in terms of cell size and time step. In this paper, we present a self-aware mesh refinement algorithm which uses a coarse mesh and a fine mesh that dynamically expands based on the plasma profile topology to resolve the sharp gradients in E-fields and plasma density in the breakdown region. The dynamic mesh refinement (DMR) technique is explained in details and its performance has been evaluated using a standard benchmark microwave breakdown problem. We observe a speedup of 8 (of the order of $O(r^{3})$, when refinement factor ($r$) is 2) compared to a traditional single uniform fine mesh based simulation. 
The technique is scalable and performs better when problem size increases. We also present a comprehensive spatio-temporal visual analysis to explain the complex physics of HPM breakdown, leading to self-organized plasma filamentation.
\end{abstract}
\section{Introduction} \label{sec1}

High power microwave (HPM) breakdown have been studied theoretically as well as experimentally since the 1950s for a wide variety of applications such as aerodynamic flow control, combustion,  precision radar systems, energy deposition in super- and hyper-sonic gas dynamic flows, beamed energy propulsion, deep-space communications applications, electromagnetic warfare, ultra-wideband (UWB) HPM transmission, to name a few ~\cite{ ADMacdonald1966,Khodataev2008,Nusinovich1997,Saifutdinov2019,MFukunari2018,JBenfordHPMdeepspapp2008,HPMwarfare2000,HPMeffectUWBtransmission2010}. 
However, only recently, detailed experimental investigations of the plasma dynamics during breakdown have been possible with the use of sophisticated high-speed cameras \cite{Hidaka2009,AMCook2011,Schaub2016,Fukunari2019,Oda2020}. Particularly, in the past few years, several experiments using millimeter and sub-millimeter wave irradiation (100s of GHz) at high pressures (ten to hundreds of Torr) have been carried out~\cite{Khodataev2008,Hidaka2008,Hidaka2009,Fukunari2019,Schaub2016,Aleksandrov2006,Oda2020,Yoda2006,AMCook2011,PBulat2021}. Different types of gas discharges have been reported experimentally such as streamer, overcritical, subcritical, volumetric and attached to an initiator~\cite{Khodataev2008,Aleksandrov2006,PBulat2021}. To completely understand the properties of each type of discharge, it is crucial to further improve our current understanding of the electromagnetic (EM) wave-plasma interaction, plasma formation, the subsequent energy exchange between wave and plasma, and
afterwards between the gas and the plasma. 
The modeling and simulation of the interaction of high power high frequency EM wave with air/gas is a computationally challenging problem \cite{BChaudhury2010,Konstantinos-2015,Saifutdinov2021,HAMIAZ2020,BCHAUDHURY2018,Kourtzanidis20143DHPM}. 
Different semi-analytical models and computational techniques have been used to study this problem~\cite{NamVerboncoeur2009,BChaudhury2010,BChaudhuryPRL2010,Nakamura2018,Saifutdinov2021,BCHAUDHURY2018,Konstantinos-2015,Kourtzanidis20143DHPM,Yan2019DGTDmesh, VGBrovkin2020,Ref2Wang2020,Bhaskarieee,HAMIAZ2020,Ref3Zhao2011}. \\
\indent
Recent experiments \cite{Hidaka2008,Hidaka2009,Aleksandrov2006, AMCook2011, Khodataev2008,Schaub2016,Yoda2006,Oda2020,Fukunari2019,MFukunari2018,PBulat2021} reveal the formation of self-organized plasma structures, either fish-bone like filaments (in overcritical), or, comb-shaped and branching (in sub-critical), that occur during air breakdown from focused (MW/cm$^{2}$), or, non-focused (kW/cm$^{-2}$) high-frequency microwave under atmospheric pressure, respectively. The plasma structures propagate towards the microwave source. The filaments elongate in the direction of the microwave electric field \cite{BChaudhury2010, Konstantinos-2015}.
The high-density plasma filaments enhance the scattered EM field at the anti-node of standing waves that result from continuous ionization diffusion mechanism and sustain the self-organized plasma pattern formation \cite{BChaudhury2010}. This problem has been used as a benchmark problem by several researchers to perform new investigations or validating the computational techniques \cite{Kourtzanidis20143DHPM,Konstantinos-2015,Semenov2015,MTakahashietal2017,Takahashi-2019,Yan2019DGTDmesh,HAMIAZ2020,Pratik2020}.\\
\indent
Accurate 2D simulations of HPM breakdown experiments mentioned above have been performed using the well established fluid model involving the coupling of Maxwell's equations and plasma continuity equation \cite{BChaudhury2010,BCHAUDHURY2018,QZHAO2011,Ref2Wang2020,Konstantinos-2015}.
It is a complex multi-physics multi-scale model due to the presence of different space and time scales \cite{Yan2019DGTDmesh, BCHAUDHURY2018} which needs to be resolved accurately. Most of the previous works \cite{Bhaskarieee,BChaudhury2010,BCHAUDHURY2018,BChaudhury2011,QZHAO2011,Konstantinos-2015,Kourtzanidis20143DHPM,Takahashi-2019,Nakamura2018,Pratik2020,HPMeffectUWBtransmission2010,Ref3Zhao2011,HAMIAZ2020} used finite difference time domain (FDTD) method in the Maxwell-plasma fluid model based simulations.
Though the simulations quite well reproduce the experimental observations but at the cost of high computation, therefore most of the past 2D numerical investigations using the fluid model have been carried out till hundreds of nanoseconds.
The excessive computational cost is due to stringent restrictions on the grid spacing and time steps \cite{BChaudhury2010,BCHAUDHURY2018,HAMIAZ2020,Konstantinos-2015}. Therefore, it is challenging to simulate large problem sizes over longer timescales (tens of microseconds) using homogeneous mesh having the finest resolution of grid size to capture the gradients in plasma density and the secondary E-field that originates from it. 
Simulation of this complex phenomenon at longer timescales will further help to understand the underlying physics for various applications in microwave rockets \cite{Takahashi-2019, MFukunari2018, Masayuki2017tf}, aerospace research \cite{Khodataev2008,DKnight2009,BCHAUDHURY2018,PBulat2021}, high-speed combustion \cite{JBMichael-2010}, safe operation of high power microwave devices ~\cite{Ref2Wang2020} etc.\\
\indent
Recently, advanced parallelization strategies for emerging many-core architectures have been proposed which significantly reduces the simulation-time, but this requires sophisticated computing facilities~\cite{BCHAUDHURY2018}.
To address the computational challenges associated with the 2D FDTD based EM-plasma fluid model for HPM breakdown, alternative numerical techniques have been developed for both structured meshes \cite{Konstantinos-2015} and unstructured meshes \cite{Yan2019DGTDmesh}. Though, the techniques are capable to balance the trade-off between accuracy and computational cost but requires complex mathematical formulation, proper choice of higher order basis functions and modification of the existing model. Recently, a static mesh refinement (MR) technique for the well established FDTD based fluid model has been presented for studying evolving plasma dynamics \cite{Pratik2020}.
Although, the static MR based technique is accurate and relatively fast compared to single uniform fine mesh, but overall performance is restricted by the size of the fixed preset refined mesh that restricts bigger and longer simulation. Therefore,  a self-aware dynamic mesh refinement (DMR) technique that generates fine mesh on demand based on the plasma evolution is proposed in this paper
The key contributions of this paper are as follows:
\begin{itemize}   
    \item Development and implementation of DMR technique for Maxwell-Plasma fluid model for investigating complex plasma dynamics during HPM breakdown.
    \item Validation and performance analysis of the proposed DMR technique against published results.
    \item To understand and visualize the plasma pattern formation and its spatio-temporal evolution under real experimental conditions using the DMR based code.
\end{itemize}

The remainder of the paper is organized as follows: Section II provides a brief introduction of the physical model, its numerical implementation and the computational challenges associated with a benchmark simulation. In section III, we discuss the theory and implementation of the Dynamic Mesh refinement technique. In Section IV, we report about the accuracy and efficiency of the proposed DMR technique, and subsequently we use our simulation results to provide a detailed spatio-temporal analysis of the complex plasma dynamics during HPM breakdown followed by conclusion in section V.
\section{Physical and Computational Model}\label{sec2}
\subsection{Physical Model}
The microwave breakdown in air/ gases at high pressure leading to complex plasma dynamics is highly collisional and nonlinear process \cite{BChaudhuryPRL2010,Konstantinos-2015,BChaudhury2010,BChaudhury2011,MTakahashietal2017,Takahashi-2019,Fukunari2019,BCHAUDHURY2018,Ref2Wang2020,Ref3Zhao2011}. 
The well established EM plasma fluid model used by several researchers \cite{BChaudhuryPRL2010, BChaudhury2010,BChaudhury2011,NamVerboncoeur2009,Ref3Zhao2011,Yan2019DGTDmesh,Ref2Wang2020,Konstantinos-2015, Kourtzanidis20143DHPM,HAMIAZ2020} to reproduce the experimental observations\cite{AMCook2011,Hidaka2008,Hidaka2009} primarily comprises solution of Maxwell's and plasma continuity equations~\cite{BChaudhuryPRL2010,BChaudhury2010}. The electron current density ($J$) couples both set of equations (Maxwell's and Plasma) \cite{BChaudhury2010}. The plasma is assumed to be quasi-neutral, and ion contribution to the current density is negligible.  
The fluid continuity equation governs the temporal evolution of the plasma density averaged over one period of the EM wave \cite{BCHAUDHURY2018}. 
Under the local electric field or effective E-field ($E_{\text{eff}}$) approximation, the gain of energy and momentum from the electric field is balanced by the losses through collisions. This equilibrium results in the local electric field to govern the ionization and attachment processes. Also, the diffusion mechanism dominates over drift for the plasma propagation in the high-frequency regime. Therefore, only the diffusive term appears in the flux divergence term in the continuity equation \cite{BCHAUDHURY2018}. The effective diffusion coefficient acts as a transition between the bulk ambipolar diffusion to free-electron diffusion in the plasma front \cite{BChaudhuryPRL2010,BChaudhury2010}.
The following four equations are primarily solved for investigating the spatio-temporal evolution of the plasma during the gas breakdown:
\begin{equation}\label{eq1}
\centering
    \frac{\partial E}{\partial t}\:=\: \frac{1}{\epsilon_0}(\boldsymbol{\nabla}\times H)\:-\: \frac{1}{\epsilon_0}\:(J)
\end{equation}
\begin{equation}\label{eq2}
\centering
    \frac{\partial H}{\partial t}\:=\: -\frac{1}{\mu_0}(\boldsymbol{\nabla}\times E)
\end{equation}
\begin{equation}\label{eq3}
\centering
    \frac{\partial v_e}{\partial t}\:=\:-\ \frac{e\ E}{m_e}\:-\:\nu_m\ v_e
\end{equation}
\begin{equation}\label{eq4}
\centering
    \frac{\partial n_e}{\partial t}-\boldsymbol{\nabla.}(D_{\text{eff}}\boldsymbol{\nabla} n_e)=n_e(\nu_i-\nu_a)-r_{ei}n_e^{2}
\end{equation}
where, $\mu_0$ and $\epsilon_0$ stands for magnetic permeability and electrical permittivity of vacuum respectively, $J$ is the plasma current density $\left(J=-e\:n_e\: v_e\right)$ in (A m$^{-2}$), $n_e$ is the plasma density (here electron density) in (m$^{-3}$), $v_e$ is the electron velocity in (m/s), $\nu_m$ is the electron-neutral collision frequency in ($s^{-1})$ (for air, $\nu_m=5.3\times10^{9}\:p$, where $p$ is the ambient pressure in (torr) \cite{BChaudhury2010}. The details related to the effective diffusion coefficient ($D_{\text{eff}}$) in (m$^{2}$/s) can be found in ~\cite{BChaudhuryPRL2010,BChaudhury2010}. In Eq (\ref{eq4}), $\nu_i\:,\:\nu_a$ are ionization and attachment frequencies respectively which is used to calculate the effective ionization, $\nu_i-\nu_a=\gamma v_d$, in (s$^{-1}$), where  $v_d$ is the electron drift velocity in (m/s). 
The ionization coefficient in terms of $E_{\text{eff}}/p$ is given by $\gamma=A p [exp(-Bp/E_{\text{eff}}]$ in (m$^{-1}$), the coefficients $A$ and $B$ correspond to air whose values are decided based on $E_{\text{eff}}/p$, in (V/cm.torr) \cite{BChaudhury2010}. The recombination coefficient($r_{ei}$) in (m$^{-3}$s$^{-1}$) is given by, $r_{ei}=\beta \times 10^{-13}(300/T_e)^{1/2}$, $\beta$ varies between $0$ and $2$. The $D_{\text{eff}}$, $\nu_i$, $\nu_a$ and $r_{ei}$ forms the transport coefficients for the plasma continuity equation\cite{Pratik2020}.
\subsection{Computational Modelling and Benchmark Problem}
\begin{figure}[!htbp]
    \centering
    \includegraphics[width=0.6\textwidth, height=0.216\textwidth]{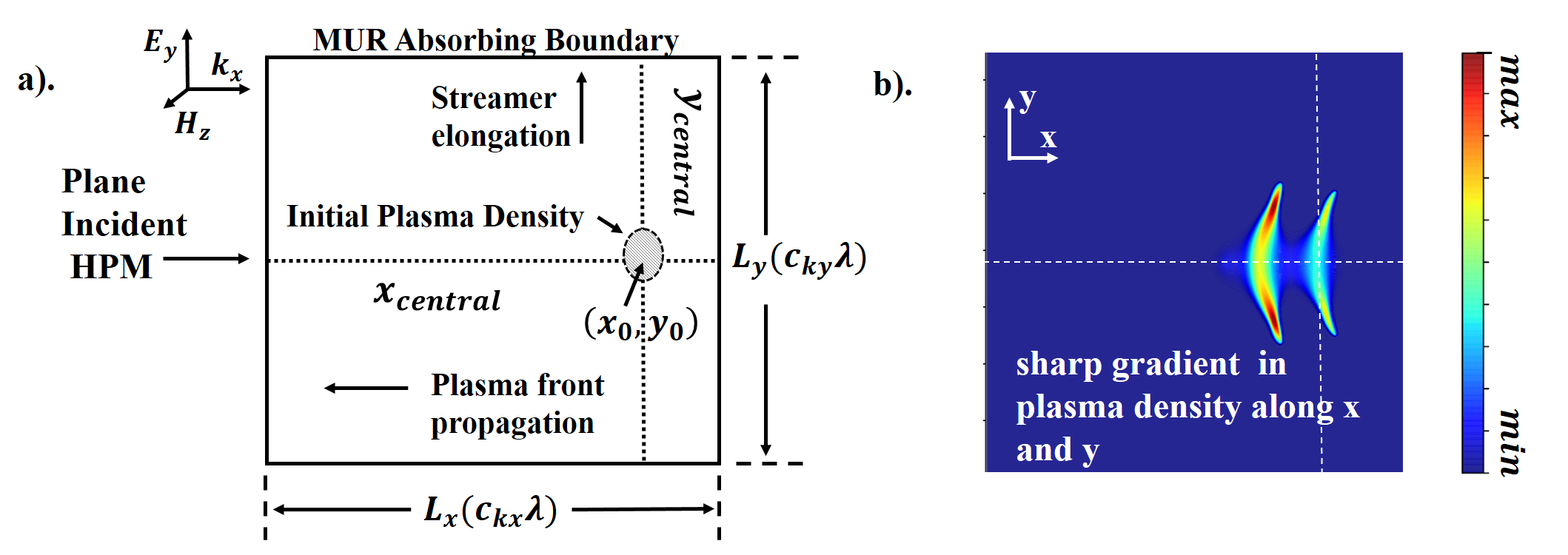}
    \caption{ (a). Schematic of the Computational domain. $\{(c_{kx},c_{ky})\in \mathbb{Q^{+}}\}$, and, $x_{0}$ and $y_{0}$ are fractions in $[0,1]$ of $L_{x}$ and $L_{y}$ respectively. The MUR outer radiation boundary condition has been used for scattered field formulation.
    (b). Formation of self-organized plasma filaments during HPM breakdown (snapshot at $t=45$ ns, $E_{0}\:=\:5.5$ MV/m, freq = 110 GHz). The maximum density ($max$) is $6\times10^{21}$ m$^{-3}$ and ($min$) is 0.
    }
    \vspace{-1mm}
    \label{fig:serialrepsen}
\end{figure}

The solution to the EM-plasma fluid model described above can be  numerically achieved by using two coupled computational solvers, EM wave solver and Plasma solver. 
We use the Yee\cite{Yee-1966} cell-based FDTD scattered field formulation to solve Maxwell's equation in the EM solver. Plasma solver uses a finite difference (FD) scheme to computationally solve the plasma continuity equation. The plasma solver requires the diffusion and growth or decay associated terms that use the effective E-field provided by the EM wave solver \cite{BChaudhury2010}. The plasma density and EM fields are evaluated at specific location on the overlapped cartesian grids \cite{BCHAUDHURY2018}. The evaluation involves two events of different timescales, the fast-evolving EM wave that requires frequent E and H-field updates with smaller time steps and the slow evolving plasma density that requires less frequent updates with bigger time step \cite{Pratik2020}. \\
\indent
Before the DMR implementation is explained in details, we describe the benchmark simulation setup on which the proposed DMR has been applied and evaluated. We consider the HPM breakdown (in air at atmospheric pressure) induced filamentary plasma propagation problem as reported in \cite{BChaudhuryPRL2010,BChaudhury2010, BCHAUDHURY2018}. Fig. \ref{fig:serialrepsen} (a), shows a schematic of our computational domain. A linearly polarized plane wave (110 GHz) propagating in air at atmospheric pressure, is incident from the left-hand side of the domain. The Electric (E) field is Y-directed in the plane of the domain, and wave propagation vector ($k$) is along X. The incident field is larger than the breakdown field. Initial plasma density with a Gaussian profile is considered at a small region centered at ($x_{0}$,$y_{0}$) which eventually evolves into a self-organized plasma filamentary structure and propagates from right to left towards the HPM source as simulation progresses \cite{Hidaka2008,BChaudhuryPRL2010,BChaudhury2010, BCHAUDHURY2018}.\\
\indent
Fig.\ref{fig:serialrepsen} (b) shows the presence of sharp gradients in the plasma density along x and y direction during the evolution of the plasma structure. For accurate simulations, $n_e$ and EM fields need to be calculated on a very fine mesh which can resolve the sharp gradients in $n_e$ and $E$~\cite{BCHAUDHURY2018, BChaudhury2011}.
The recommended minimum number of grid points per wavelength ($N_{\lambda}$) of the electromagnetic (EM) wave is around 500, i.e. $N_{\lambda}\:=\:\lambda/\Delta >500$, for E-field in simulation plane ~\cite{BCHAUDHURY2018}. Here, $\Delta$ is the grid size. Requirement of very fine grids significantly increases the computational cost (in terms of number of grid points in the simulation domain and a very small time step) for high frequency problems such as 100's of GHz ($\lambda\:\approx\:2$ mm). Since the simulation follows an iterative algorithm, the total simulation time increases with the increase in problem size and decrease in time step.
 This makes it very challenging to handle bigger problem sizes in 2D simulations.
On a latest standard desktop, a well accepted benchmark simulation of a $1\lambda\times1\lambda$ problem using uniform mesh, $N_{\lambda}\:=\:512$, takes around 5 days. 
The computational complexity scales in the order of $O(N_{f})$, here $N_{f}=c_{kx}c_{ky} N^{2}_{\lambda}$, is total cells in the mesh, as $c_{k}$ changes with problem size. Therefore, for simulating a $10\lambda\times10\lambda$ ($c^{2}_{k}=(10)^{2}$) problem using a serial code on a standard desktop it will take around 500 days. This possesses a real computational challenge.
\section{Self Aware Mesh Refinement Technique}\label{sec3}
\begin{figure}[!htbp]
    \centering
    \includegraphics[width=0.7\textwidth, height=0.182\textwidth]{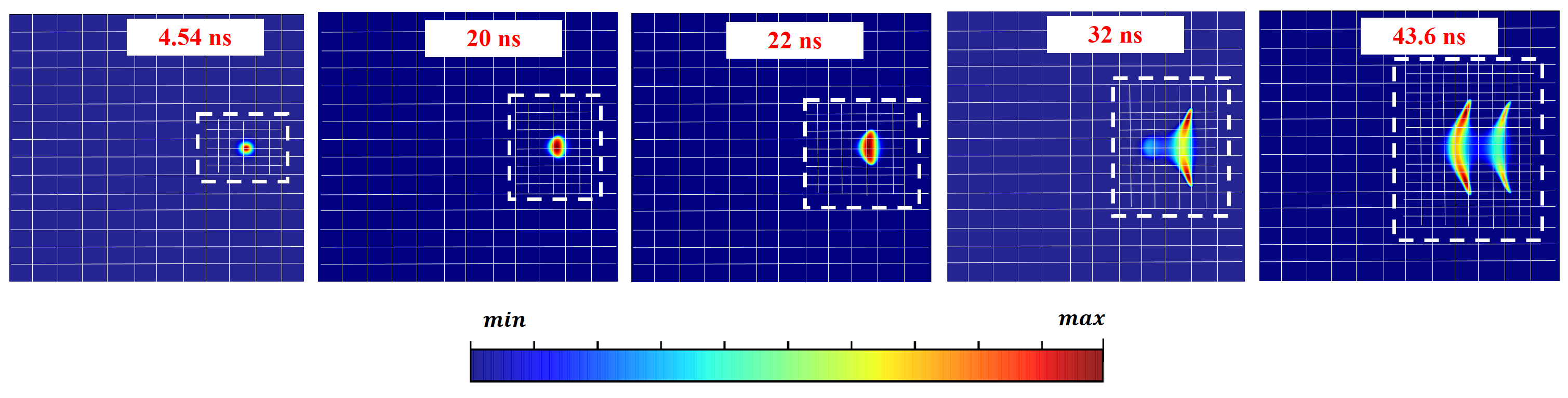}
    \caption{ The self-aware expansion of fine mesh in DMR to capture the 2D distribution of plasma density as filamentary pattern evolves.
    }
    \vspace{-3mm}
    \label{fig:dmrrepresen}
\end{figure}

The proposed self-aware mesh refinement technique or DMR technique is developed on the basic framework of static Mesh Refinement (MR). MR hierarchically decomposes the computational domain, into coarse and fine mesh, that are overlapped and logically connected to maintain the continuity between the evaluated quantities on both mesh, \cite{Zivanovich1991,KaiXiao2007,CostasSaris-2007,Pratik2020}. In case of static MR, the fine mesh region remains static/preset along with the coarse mesh.\\
\indent
The conventional MR (static/preset fine mesh) can not be used very efficiently for a system which is evolving in time and space as in the case of plasma evolution during HPM breakdown. Also, if the preset fixed refinement region is large, then it leads to unnecessary large computations. The proposed DMR technique considers a fixed coarse mesh and an expanding fine mesh based on the evolving plasma profile.
Fig. \ref{fig:dmrrepresen} shows how the initial fine mesh (confined within a small fraction of total computational domain) expands in the proposed DMR technique. The mesh expansion varies the amount of refinement region as a function of time ($t$), denoted by $R(t)$.
Fine mesh expansion procedure is explained below in details.
\subsection{Initiation of fine mesh expansion}
The initiation of fine mesh expansion requires detection of sharp variations in evaluated quantities (E and H-fields, plasma density etc.) on the discretized grids and use a suitable fine mesh to capture such variations. 
Generally, to determine the grid size, threshold criteria based on gradients in overall and instantaneous energy \cite{CostasSaris-2007} are used widely. The criteria hold true for EM-scattering problems, where improper resolution of scattering geometry results in higher scattered E-fields. 
For the HPM breakdown, when the medium properties are continuously evolving, we adopt alternative criteria that decides the initiation of mesh expansion and the size of grid to resolve sharp variations.\\
\begin{figure}[!htbp]
    \centering
    \includegraphics[width=0.5\textwidth, height=0.225\textwidth]{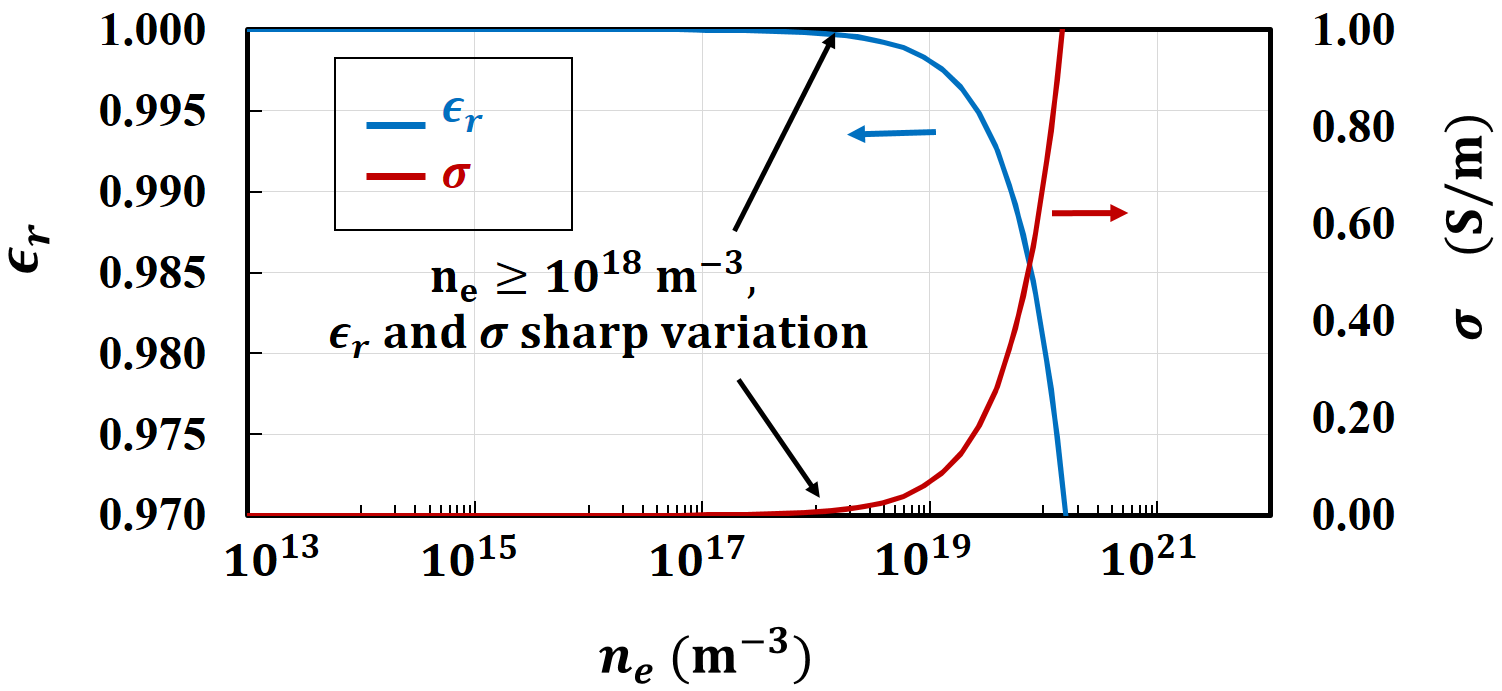}
    \caption{ The spatial variation of $\epsilon_r$ and $\sigma$ for a 1D plasma density ($n_{e}$) distribution. The EM wave of frequency 110 GHz is considered here.
    }
    \vspace{-2mm}
    \label{fig:dmrmeshtrigring}
\end{figure}
\indent
The complex relative permittivity for a collisional plasma is defined as \cite{BChaudhury2010,BChaudhury2011},
\begin{equation}
    \epsilon_r=\left(1-\frac{\omega_{pe}^{2}}{\omega^2+\nu^{2}_{m}}\right)-i\left(\frac{\omega_{pe}^{2}}{\omega^2+\nu^{2}_{m}}\right)\left( \frac{\nu_{m}}{\omega}\right)
\end{equation}
where $\omega_{pe}= \left(n_e e^{2}/m \epsilon_0\right)^{1/2}$ is electron plasma frequency. The $Re{\{\epsilon_r\}}$ decides the propagation of the electromagnetic wave in the plasma, whereas the conductivity of the plasma depends on $Im{\{\epsilon_r\}}$. 
The cutoff density ($n_{\text{cutoff}}$), in the context of EM propagation in plasma, is given by, $ n_{\text{cutoff}}=n_{\text{critical}}\left(\nu_m/\omega\right)$, where $n_{\text{critical}}= (m \epsilon_0/e^{2})\omega$. 
When plasma density crosses the cutoff density, plasma starts reflecting the EM wave \cite{BChaudhury2010}.\\ 
The decision to initiate the fine mesh expansion depends on the occurrence of a fixed plasma density (we refer to it as threshold density). The threshold density criteria are well validated based on EM propagation characteristics (or dispersion relation) in plasma \cite{BHASKAR2007}. Threshold density in our method is determined by performing a convergence study described in next section, when the occurrence of minimum variation in permittivity ($\epsilon$) and conductivity ($\sigma$) occurs for a given plasma density profile and incident EM wave frequency, as observed in Fig. \ref{fig:dmrmeshtrigring}. For deciding the grid size of coarse and fine mesh, we take an inverse approach where the finest grid size is pre-decided based on valid assumptions \cite{BCHAUDHURY2018} and the coarse grid size is decided based on the mesh refinement factor. The cell size remains fixed for both the fine mesh and the coarse mesh during the simulation.  
\begin{figure*}[ht]
    \centering
    \includegraphics[width=\textwidth, height=0.5\textwidth]{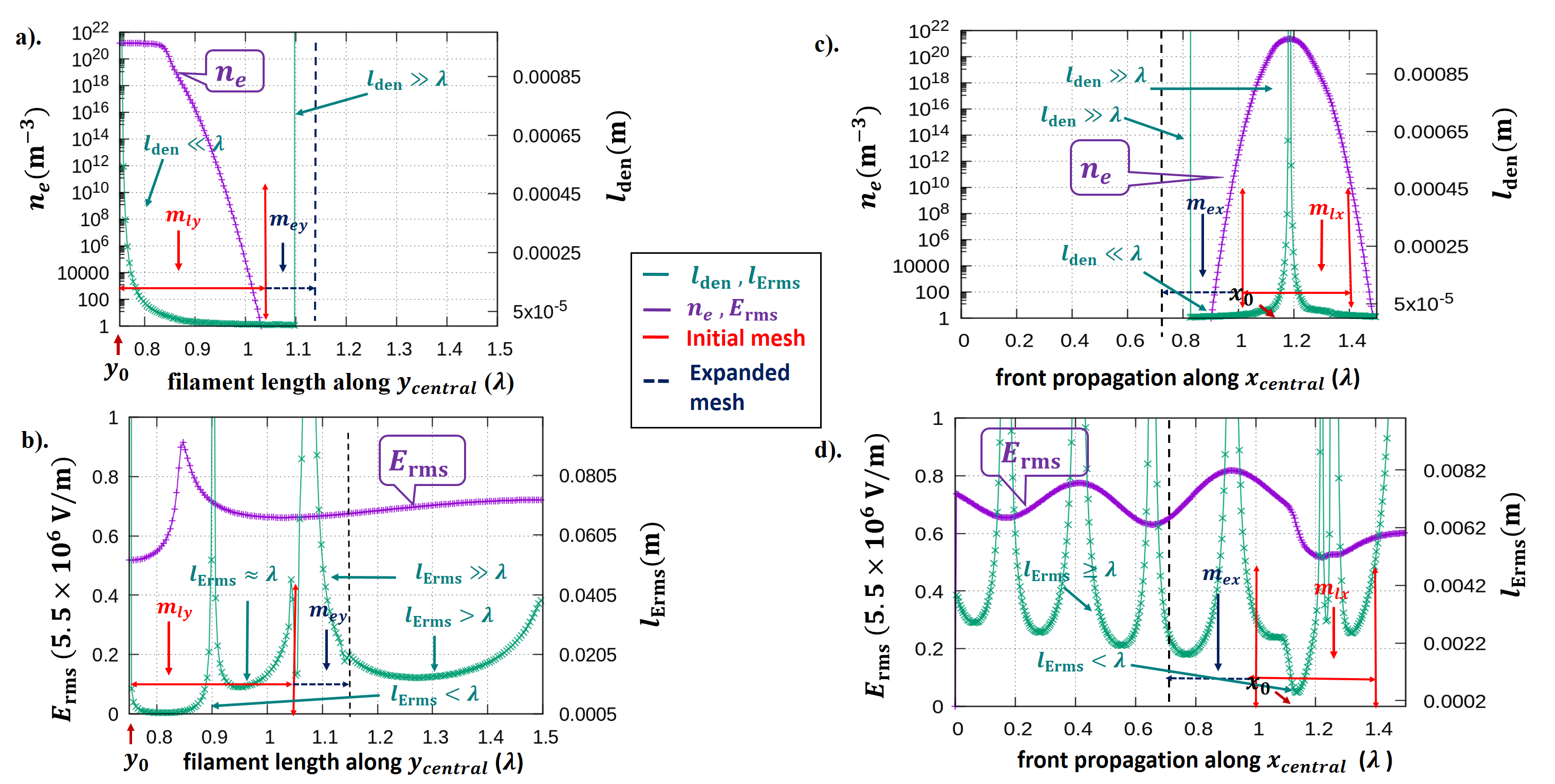}
    \caption{ The 1D distribution of plasma density (a) and rms E-field (b) and their corresponding gradient scale length $l_{\text{den}}$ and $l_{E_{\text{rms}}}$ respectively, along upper half of central y-axis ($y_{central}$) through initial plasma density. Similarly, in (c,d) along the central x-axis ($x_{central}$) through the initial plasma density. The initial plasma density is located at $\{x_{0},y_{0}\}\:=\:\{0.85L_{x},0.5L_{y}\}\:\:=\{1.2\lambda,0.75\lambda\}$, $L_{x}\:=\:L_{y}\:=\:1.5\lambda$. $m_{lx}$ and $m_{ly}$, and, $m_{ex}$ and $m_{ey}$ are the length of the initial refinement region and the mesh expansion along x and y, respectively. 
    The $\lambda\approx0.0027$ m, correspond to frequency ($f$) = 110 GHz.
    }
    \vspace{-2mm}
    \label{fig:dmrmeshexpcrit}
\end{figure*}
\subsection{Amount of fine mesh expansion}
\indent
Fig. \ref{fig:serialrepsen} (b) shows the presence of sharp gradients in the plasma density during HPM breakdown and is an important parameter to decide the amount of fine mesh expansion. 
We have considered two gradient length scales corresponding to plasma density and rms E-field (scattered field from plasma), $l_{\text{den}}$ and $l_{E_{\text{rms}}}$, respectively. Mathematically, $l_{\text{den}}\:=\: n_{e}/|\nabla n_{e}|$ and $l_{E_{\text{rms}}}\:=\: E_{\text{rms}}/|\nabla E_{\text{rms}}|$, where $|\nabla n_{e}|$ and $|\nabla E_{\text{rms}}|$ are magnitudes of gradient in density and rms E-field respectively.\\
\indent
For the expansion of the fine mesh, the growth of filaments and the associated density gradients along $y_{central}$ and $x_{central}$ as shown in Fig. \ref{fig:serialrepsen} (a-b) have been considered. Due to symmetric nature of the plasma propagation, the region above $x_{central}$ starting at $y_o$ is only considered for analysis.
In Fig. \ref{fig:dmrmeshexpcrit} (a-d), the length of initial fine mesh, and, the amount of fine mesh expansion along x and y are represented by, $m_{lx}$ and $m_{ly}$, and, $m_{ex}$ and $m_{ey}$, respectively. Fig. \ref{fig:dmrmeshexpcrit} (a,b) shows the distribution of plasma density, rms E-field and the corresponding gradient length scales along y-direction.
The $l_{\text{den}}$ has a sharp transition from a high value, $l_{\text{den}}>>\lambda$ ($\nabla n_{e}\to 0$), to a low value, $l_{\text{den}}\:<\: (1/100) \lambda$. 
There is a high gradient in rms E-field that exists from $0.75\lambda$ to $0.85\lambda$ see Fig. \ref{fig:dmrmeshexpcrit} (b), and as a result the gradient length scale ($l_{E_{\text{rms}}}$) transits from low ($l_{E_{\text{rms}}}<\lambda\sim (1/5) \lambda$) to high ($l_{E_{\text{rms}}}>>\lambda$, when $(\nabla E_{\text{rms}}\to 0)$). The initial fine mesh centered around initial density can capture the sharp gradients in both, density and rms E-field along $y_{central}$ provided, $m_{ly}$ is large enough to cover the occurrence of smallest, $l_{den}$ as well as $l_{E_{\text{rms}}}$, and the fine mesh grid size satisfy, $\Delta S\sim min\:\{l_{\text{den}},l_{E_{\text{rms}}}\}$. Next, for the fine mesh expansion along $y_{central}$, the amount of expansion, $m_{ey}$, must be able to capture the occurrence of smallest, $l_{\text{den}}$ as well as $l_{E_{\text{rms}}}$, as specified above.\\
\indent
Similarly, Fig. \ref{fig:dmrmeshexpcrit} (c,d) presents the distribution of plasma density and rms E-field and their corresponding gradient length scales in the x-direction (along $x_{central}$; the plasma front propagation direction). The amount of mesh expansion is determined by the presence of gradient length scales.
By referring Fig. \ref{fig:dmrmeshexpcrit} (a-d), it can be observed that $min\{l_{\text{den}}\} < min\{l_{E_{\text{rms}}}\}$. Thus, $l_{\text{den}}$, instead of $l_{E_{\text{rms}}}$, decides fine mesh grid size. Based on the gradients shown in Fig. \ref{fig:dmrmeshexpcrit} (a-d), $m_{ex}\:>\:m_{ey}$. Therefore, different amount of mesh expansion is required in x and y directions.
\subsection{Quantity Updates on Mesh and Synchronization}
\begin{figure}[!htbp]
   \centering
    \includegraphics[width=0.8\textwidth, height=0.4\textwidth]{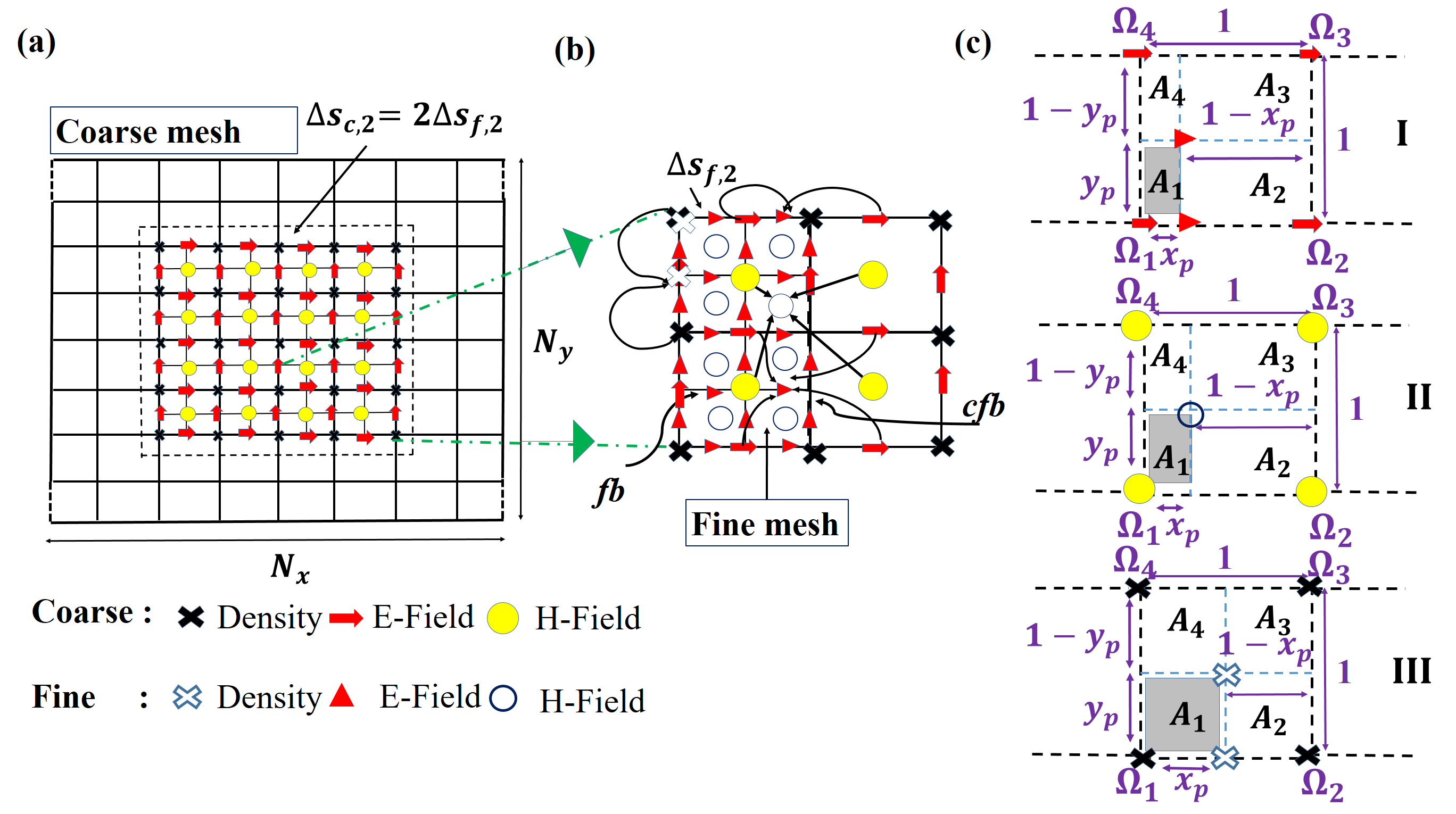}
    \caption{(a). The mesh refined (dashed region with refinement factor 2) discretized computational grid showing locations for computation of EM fields and density. 
     (b). The expanded view of overlapped coarse and fine mesh. The different data transfer of updated \textbf{E}-field, \textbf{H}-field and the plasma density from coarse mesh to its corresponding fine mesh locations on both coarse-fine boundary ($cfb$) as well as the fine boundary ($fb$) are shown. (c).Schematic representation of interpolation techniques, for r=2. The locations, $\{x_{p},y_{p}\}$ are as follows: for, $\text{I}:$ $\{1/2r,1/r\}$, $\text{II}:$ $\{1/2r,1/r\}$ and $\text{III}:$ $\{1/r,1/r\}$. Here, $\text{I}:$ E-field (and velocity), $\text{II}:$ H-field and $\text{III}:$ plasma density, represent different interpolation for the coarse mesh data.
    }
    \label{fig:parallelscheme}
\end{figure}
The schematic in Fig. \ref{fig:parallelscheme} (a) represents the mesh refinement region (in dashed). Here, a single level of refinement is shown, having overlapped coarse and fine mesh, with a grid refinement factor ($r$) of two, such that the ratio of coarse to fine grid size is $2:1$. 
As depicted in Fig.\ref{fig:parallelscheme} (b), two meshes are overlapped, coarse and embedded fine. 
E, H, density and velocity (synchronized with E) as shown in Fig. \ref{fig:parallelscheme} are updated as discussed in \cite{BCHAUDHURY2018}. 
The fields and density are updated simultaneously both on the coarse and on fine mesh maintaining the proper sequence of frequent FDTD and less frequent (on each period of EM wave) update of plasma continuity equation \cite{BCHAUDHURY2018}. The different time steps are based on respective CFL conditions. Time step associated with EM wave solver is much smaller compared to plasma solver~\cite{BCHAUDHURY2018, BChaudhury2010}.\\
The updates must be synchronized both in space and time between the two meshes to maintain continuity.
Let the grid size, time step, total cells and total iterations for benchmark case (uniform fine mesh) are represented by $\Delta S_{f}$, $\Delta t_{f}$, $N_{f}$ and $I_{f}$, respectively. In case of DMR, the coarse mesh cell size and time-step are represented by $\Delta S_{c}$ and $\Delta t_{c}$, and for fine mesh it is represented by $\Delta S_{f}$ and $\Delta t_{f}$. $\Delta t_{c}$ and $\Delta t_{f}$ are associated with Maxwell's updates (which primarily determines the total execution time). The subscripts $c$ and $f$, corresponds to coarse and fine meshes, respectively.
Here, $\Delta S_{f}=\lambda/N_{\lambda}$ and $\Delta S_{c}\:=\: r \Delta S_{f}$. Similarly, $\Delta t_{c}$=$r\Delta t_{f}$. In one coarse mesh update, $\Delta t_{c}$, the fine mesh performs $r$ updates with $\Delta t_{f}$. This follows for all the quantities.
The total fine mesh cells depend on the amount of refinement region ($R(t)$), here $R(t)$ is a fraction of total computation domain and $R(t)\in[0,1]$.\\
\indent
The MR algorithm transfers the evaluated quantities (fields, velocity and plasma density) from the coarse mesh to both, the coarse-fine boundary ($cfb$) and the fine boundary ($fb$) for subsequent fine mesh updates. This transfer of evaluated quantities supports the nearest neighborhood dependence of FDTD and FD based fields and density updates, respectively \cite{BChaudhury2010}. Finally, the fine mesh updated quantities are transferred back to the coarse mesh. Both the data transfer process must occur within the coarse mesh update interval. The process avoids discontinuity in the obtained results due to mismatch between coarse and fine mesh values. The two boundaries ($fb$ and $cfb$) and the sub-grids are shown in Fig. \ref{fig:parallelscheme} (b).\\
\indent
The data transfer uses either direct copy or interpolation process, depending on the location (coinciding or non-coinciding) of the quantities on the overlapped grids. The interpolation is either a linear interpolation on $cfb$ or a bi-linear interpolation on $fb$ as indicated by the direction of arrows in Fig.\ref{fig:parallelscheme} (b). 
Quantities, $\mathrm{\Omega_{i}}$, i=1,2,3 and 4, shown in Fig. \ref{fig:parallelscheme} (c),
represents either E-field (and velocity), H-field or plasma density on the coarse mesh. 
The dotted square represents the interpolation domain.  The vertices represent the coarse data and the desired fine data, either on the edge (1D) or inside the bounded area (2D plane), 
can be obtained using the equation as follows:
\begin{subequations}\label{eq:5}
\begin{equation}
\begin{split}
  BL_{f_{new}} &= 
   (1-frac)(\Omega_{1_{old}}A_{3} + \Omega_{2_{old}}A_{4} \\
   &\phantom{{}=} + \Omega_{3_{old}} A_{1} + \Omega_{4_{old}} A_{2}) \\
   &\phantom{{}=} + (frac)(\Omega_{1_{new}} A_{3} + \Omega_{2_{new}} A_{4} \\
   &\phantom{{}=} + \Omega_{3_{new}} A_{1} + \Omega_{4_{new}} A_{2}) \\
 \end{split}
\end{equation}
\begin{equation}
\begin{split}
  L_{f_{new}} &= 
   (1-frac)(\Omega_{1_{old}} (1-x_{p}) + \Omega_{2_{old}} (x_{p})) \\
   &\phantom{{}=} + (frac)(\Omega_{1_{new}}  (1-x_{p}) + \Omega_{2_{new}}  (x_{p}) )
\end{split}
\end{equation}
\end{subequations}
where, suffix $old$ and $new$, denotes the previous and updated quantities on coarse ($c$) mesh that requires interpolation on fine ($f$) mesh. $A_{i}$, i=1,2,3 and 4, represents the area inside the interpolation domain (dotted square) calculated in terms of $x_{p}$ and $y_{p}$ represented in terms of $r$. 
The $frac$ and $(1-frac)$ are the ratios in which the old and the updated coarse mesh data values must be taken to obtain a smoothed interpolated quantity on the fine mesh. $BL_{f_{new}}$ and $L_{f_{new}}$, represents the bi-linear and linear interpolated fine mesh data, respectively from coarse mesh data (E or H-field or velocity or plasma density).
\indent
\begin{figure}[!htbp]
    \centering
    \includegraphics[width=0.5\textwidth, height=0.325\textwidth]{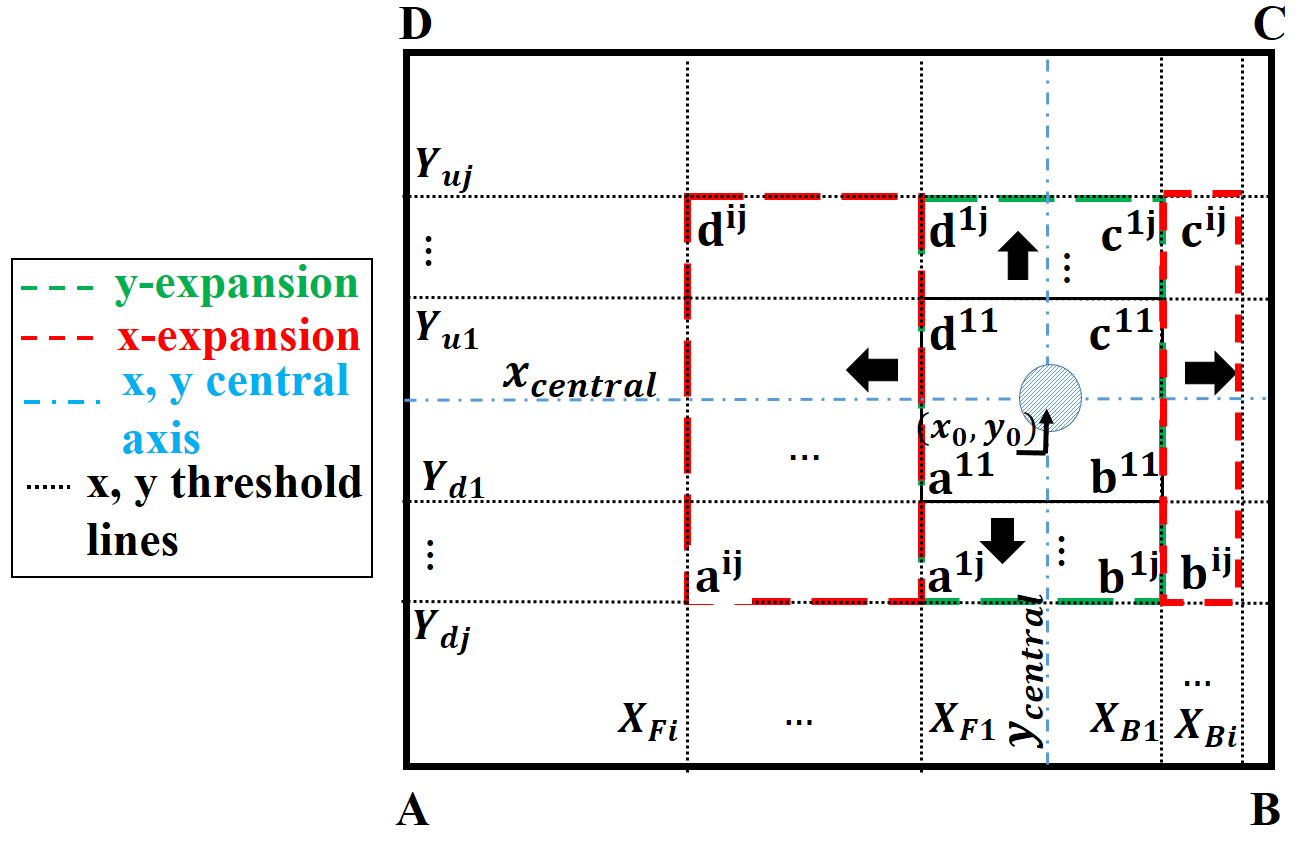}
    \caption{ Schematic representation of the dynamic mesh generation process with an initial fine mesh, $\text{a}^{11}$-$\text{b}^{11}$-$\text{c}^{11}$-$\text{d}^{11}$, centered around the initial plasma density located at $x_0$ and $y_0$, which expands along x and y based on threshold criteria. 
    The coarse mesh is present in the complete computational domain, A-B-C-D.
    }
    \vspace{-2mm}
    \label{fig:DMRexpansion}
\end{figure}
\subsection{Implementation of DMR algorithm}
First, an initial fine mesh region, $\text{a}^{11}$-$\text{b}^{11}$-$\text{c}^{11}$-$\text{d}^{11}$, shown in Fig. \ref{fig:DMRexpansion}, is considered as indicated by the black solid lines. It is located inside coarse mesh region, that covers overall computational domain, A-B-C-D. The initial fine mesh contains the initial plasma density profile located at $x_0$ and $y_0$. As discussed in the previous section, the E-field, H-field, $v_{e}$ and $n_{e}$ are updated in both the coarse mesh and the initial fine mesh. 
Next, the fine mesh generation proceeds with two steps, the self-aware initiation and expansion of initial fine mesh. Different threshold lines parallel to $x_{central}$ and $y_{central}$, indicated by forward threshold ($X_{Fi}$), backward threshold ($X_{Bi}$), upper threshold ($Y_{uj}$) and lower threshold ($Y_{dj}$), where i=1,2...n and j=1,2,...,m, where $\{(n,m)\in \mathbb{N}\}$, are considered as shown by dotted lines. The threshold lines coincide with the respective x and y boundaries of the fine mesh (happens to be the $cfb$). Before the fine mesh expansion initiates, it is checked whether $n_{e}$ is greater than the threshold density, on either, $X_{Fi}$ and $X_{Bi}$, or, $Y_{uj}$ and $Y_{dj}$. 
Based on whichever threshold line meets the threshold density criteria, the fine mesh is expanded along $x_{central}$ or $y_{central}$ resulting into a y-expanded: a$^{\text{1j}}$-b$^{\text{1j}}$-c$^{\text{1j}}$-d$^{\text{1j}}$ or  x-expanded: a$^{\text{ij}}$-b$^{\text{ij}}$-c$^{\text{ij}}$-d$^{\text{ij}}$ fine mesh region, shown in Fig.\ref{fig:DMRexpansion}.
During fine mesh expansion, first the E-, H-field $v_{e}$ and $n_{e}$ data on the entire initial fine mesh are transferred to the similar locations on the expanded fine mesh to maintain the continuity from the initial mesh. Next, for the remaining regions in the expanded fine mesh, the entire overlapped coarse data are interpolated on the fine mesh. 
The mesh expansion continues as and when required in a self-aware manner based on spatio-temporal evolution of plasma.
\setlength{\textfloatsep}{10pt plus 1.0pt minus 2.0pt}
\section{Results and Discussion}\label{sec4}
For performance analysis, we consider the same computational setup as described in Fig. \ref{fig:serialrepsen} (a), where the size of the computational domain is represented by, $L_{x}\:=\: c_{kx}\lambda$ and $L_{y}\:=\:c_{ky}\lambda$. We have taken different $c_{kx}$ and $c_{ky}$ for different computational experiments. Initial 2D Gaussian plasma density is $n_{e}(x,y)=n_{0}exp(-(\{x-x_{0}\}^{2}/\sigma_{x}^2+\{y-y_{0}\}^{2}/\sigma_{y}^2))$, where $x_{0}$ and $y_{0}$ is the location of $n_0$ and, $\sigma_{x}$ and $\sigma_{y}$, controls plasma width along x and y, respectively. We consider, $n_{0}=10^{16}$ m$^{-3}$, the Incident E-field, $E_{0}\:=\:5.5$ MV/m and frequency ($f$) =$110$ GHz. All the computations are carried out on a computer with Intel Xeon CPU E5-2640 processor with 32 GB RAM. 
\subsection{Threshold density for mesh expansion}
\begin{figure}[!htbp]
    \centering
    \includegraphics[width=0.5\textwidth, height=0.2875\textwidth]{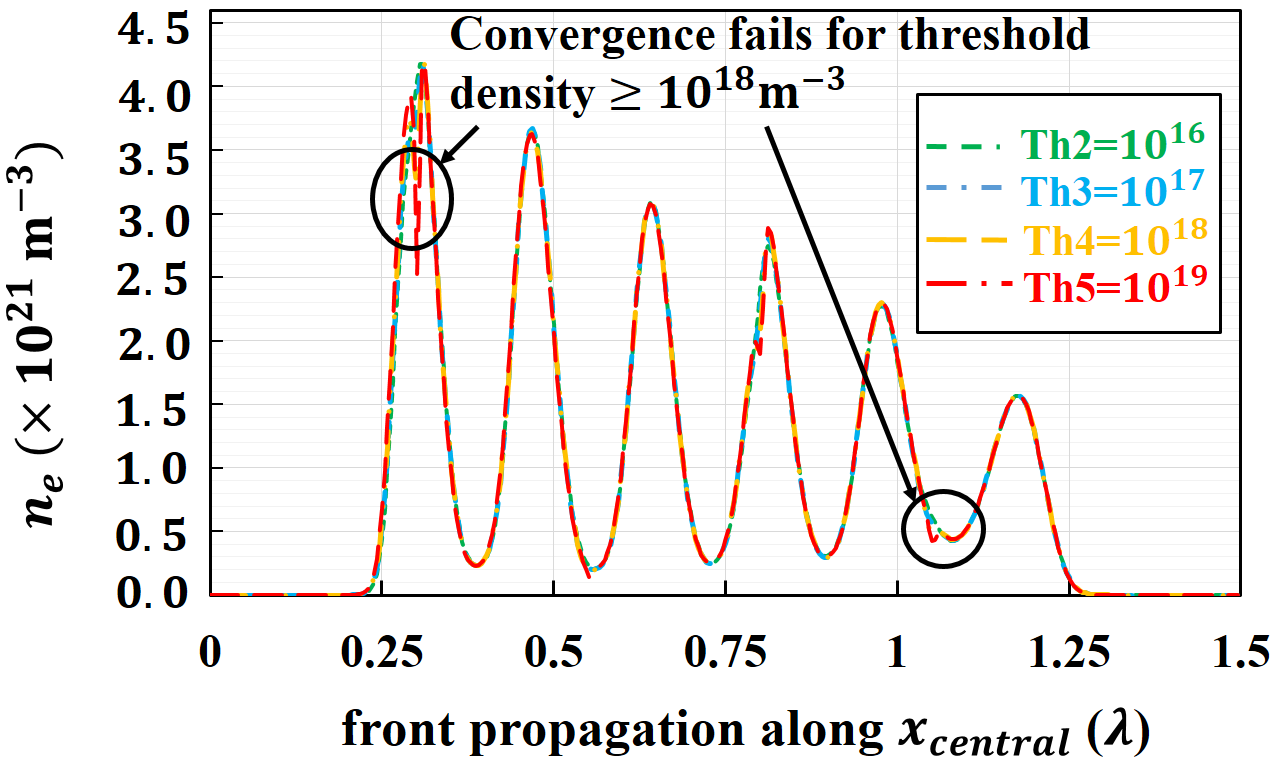}
    \caption{Convergence study for different threshold densities ($m^{-3}$) to arrive at the threshold density criteria for mesh initiation.
    }
   \vspace{-2mm}
    \label{fig:dmrmeshtrigrigval2}
\end{figure}

The optimal choice of threshold density is obtained through a convergence study for a 1D plasma density distribution along $x_{central}$ for different threshold densities, as shown in Fig. \ref{fig:dmrmeshtrigrigval2}. The failure in the convergence, for density $\geq 10^{18}$ m$^{-3}$, can be observed by the presence of sudden spikes in plasma distribution along $x_{central}$ in Fig. \ref{fig:dmrmeshtrigrigval2} (highlighted using circles).
 The optimal chosen density, $n_{e}<10^{18}$ m$^{-3}$ $\approx 10^{16}-10^{17}$ m$^{-3}$, where, both the values of $\epsilon$ and $\sigma$ have minimum variations as shown in Fig. \ref{fig:dmrmeshtrigring} satisfies lower scattered E-field and avoids sharp gradient in E-field or energy. 
The amount of mesh expansion is arrived at by referring to Fig. \ref{fig:dmrmeshexpcrit}, 
to ensure that the fine mesh resolves the minimum gradient scale lengths along y and x-axis, respectively.
\subsection{DMR Accuracy}
\begin{figure}[!htbp]
    \centering
    \includegraphics[width=0.6\textwidth, height=0.48\textwidth]{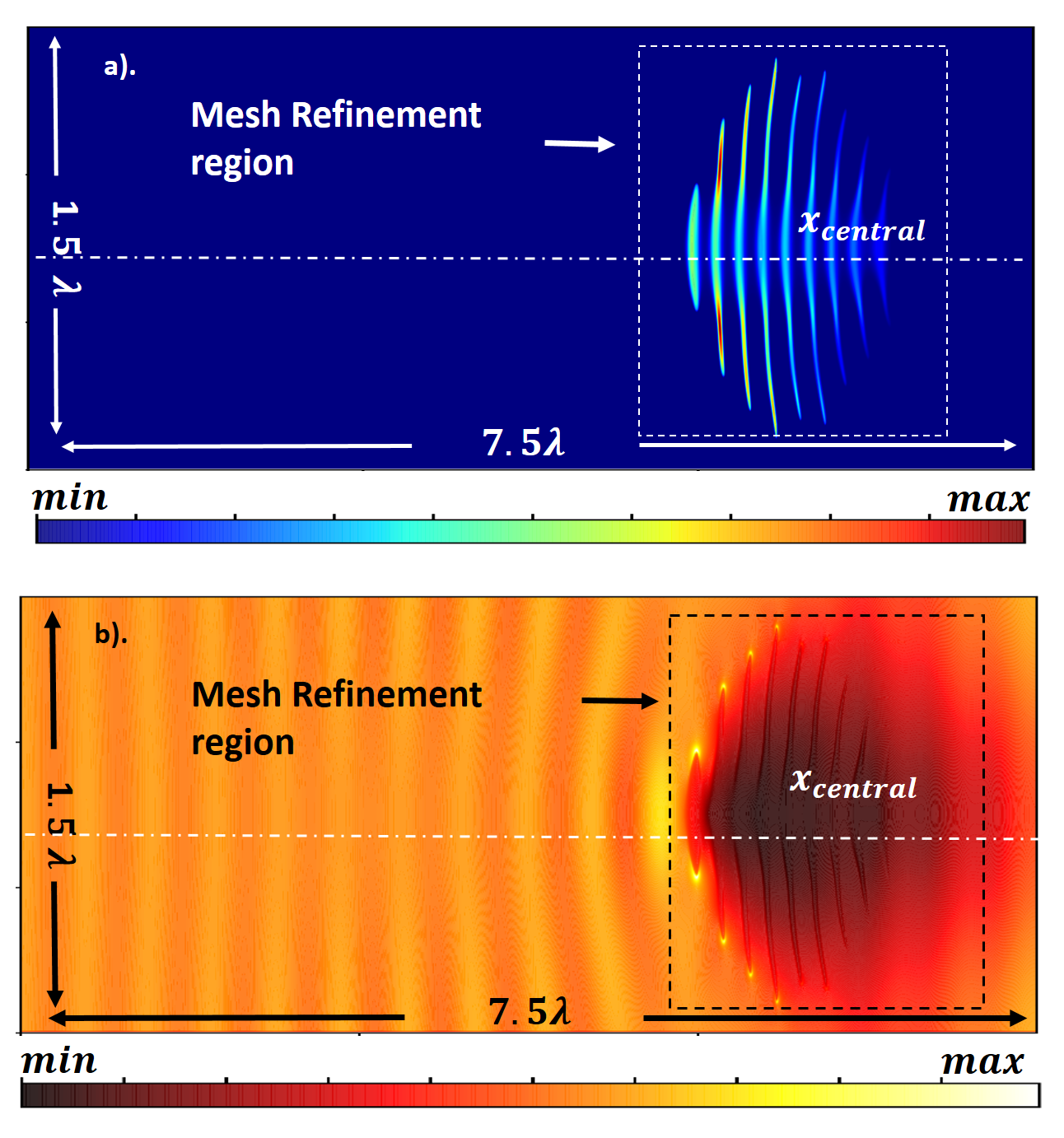}
    \caption{The 2D distribution of (a) plasma density (m$^{-3}$) at time $t=140$ ns and (b) corresponding RMS \textbf{E}-field, for a problem size of $7.5\lambda\times1.5\lambda$. The maximum density is $8.7\times10^{21}$ m$^{-3}$ and maximum \textbf{E}-field strength is $6.97\times10^{6}$ V/m as represented on the color scale. 
    }
    \vspace{-2mm}
    \label{fig:Accuracytest1}
\end{figure}

\begin{figure}[!htbp]
\centering
\includegraphics[width=0.7 \textwidth,height=0.49\textwidth]{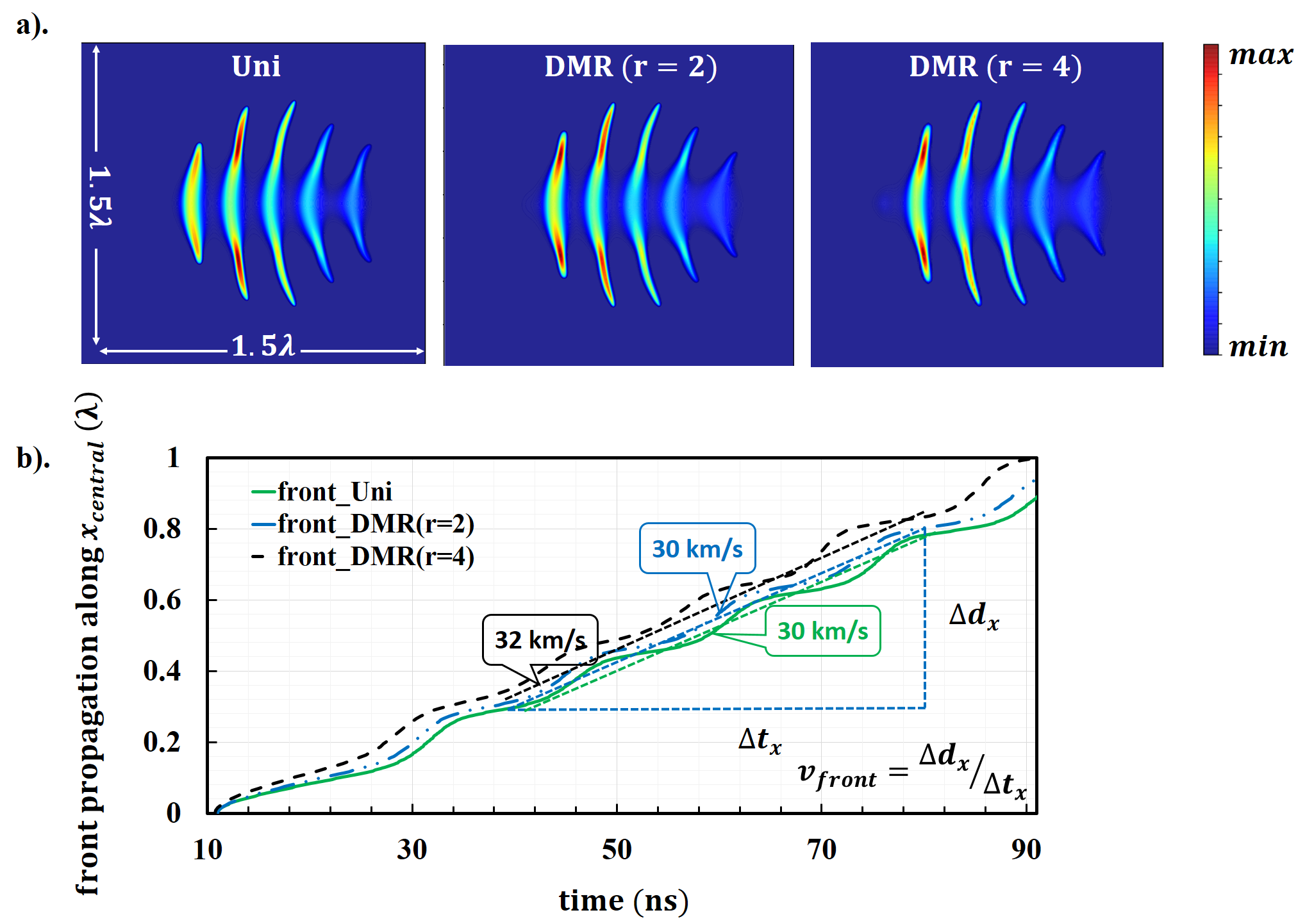}
\caption{(a).The 2D distribution of plasma density (m$^{-3}$) at time $t=90$ ns obtained using single uniform mesh (Uni), and DMR (r=2 and r=4). (b). The comparison between the plasma front propagation using temporal evolution of the plasma density (m$^{-3}$) along $x_{central}$ for uniform mesh (Uni) and dynamic mesh (DMR) with different refinement factors ($r$), here $r\:=\:2\:,\:4$. The colorbar represents plasma density distribution, $max:7\times10^{21}$ m$^{-3}$. 
}
\vspace{-2mm}
\label{accuracytest2}
\end{figure}

For evaluating the accuracy of the DMR method, the consistency of shape and size of plasma filaments, and plasma front velocity are compared with the uniform fine-mesh implementation. We consider, $\{ c_{kx},c_{ky}\}\:=\:\{7.5,1.5\}$ to investigate the size and shape of filaments.
Initial Gaussian peak plasma density ($n_{0}$) is located at $x_{0}$ and $y_{0}$, which is given by $0.85\:L_{x}$ and $0.5\:L_{y}$ respectively.
Fig. \ref{fig:Accuracytest1} (a,b), represents the distribution of plasma density and the corresponding scattered rms E-field at time $t\:=\:140$ ns, obtained using DMR technique with refinement factor ($r\:=\:2$). The  observed results are in good agreement with the published results from \cite{Hidaka2008,BChaudhury2010,BCHAUDHURY2018}. 
The dynamic mesh could capture the nonuniform gradients in both density and scattered E-field by utilizing an optimal fine mesh region shown by dotted lines in Fig. \ref{fig:Accuracytest1} (a-b). 
To validate the plasma front velocity, $\{c_{kx},c_{ky}\}\:=\:\{1.5,1.5\}$ is considered for the simulation setup. Fig. \ref{accuracytest2}, represents the temporal evolution of the plasma density along the central x-axis ($x_{central}$) in the computational domain for uniform mesh (Uni) and DMR ($r\:=\:2\:,\:4$). We can calculate the front velocity along $x_{central}$ by tracking the propagation of a specific plasma density level at the front (we have used $\approx1\times10^{19}$ (m$^{-3}$)) with time. 
The calculated plasma front velocity  ($v_{\text{front}}$) is $\approx 30 $ km/s for both uniform mesh and DMR with $r\:=\:2$. 
We found the variation in front velocity and length of filament lies within $1-2\%$ of the uniform fine mesh case (Uni). We observe the velocities and lengths are consistent with previously published experimental and simulated results \cite{Hidaka2008,Hidaka2009,BChaudhury2010,BChaudhury2011,Bhaskarieee,BCHAUDHURY2018}.
\subsection{Speedup and Efficiency of DMR technique}
\begin{figure*}[ht]
\centering
\includegraphics[width= \textwidth,height=0.57\textwidth]{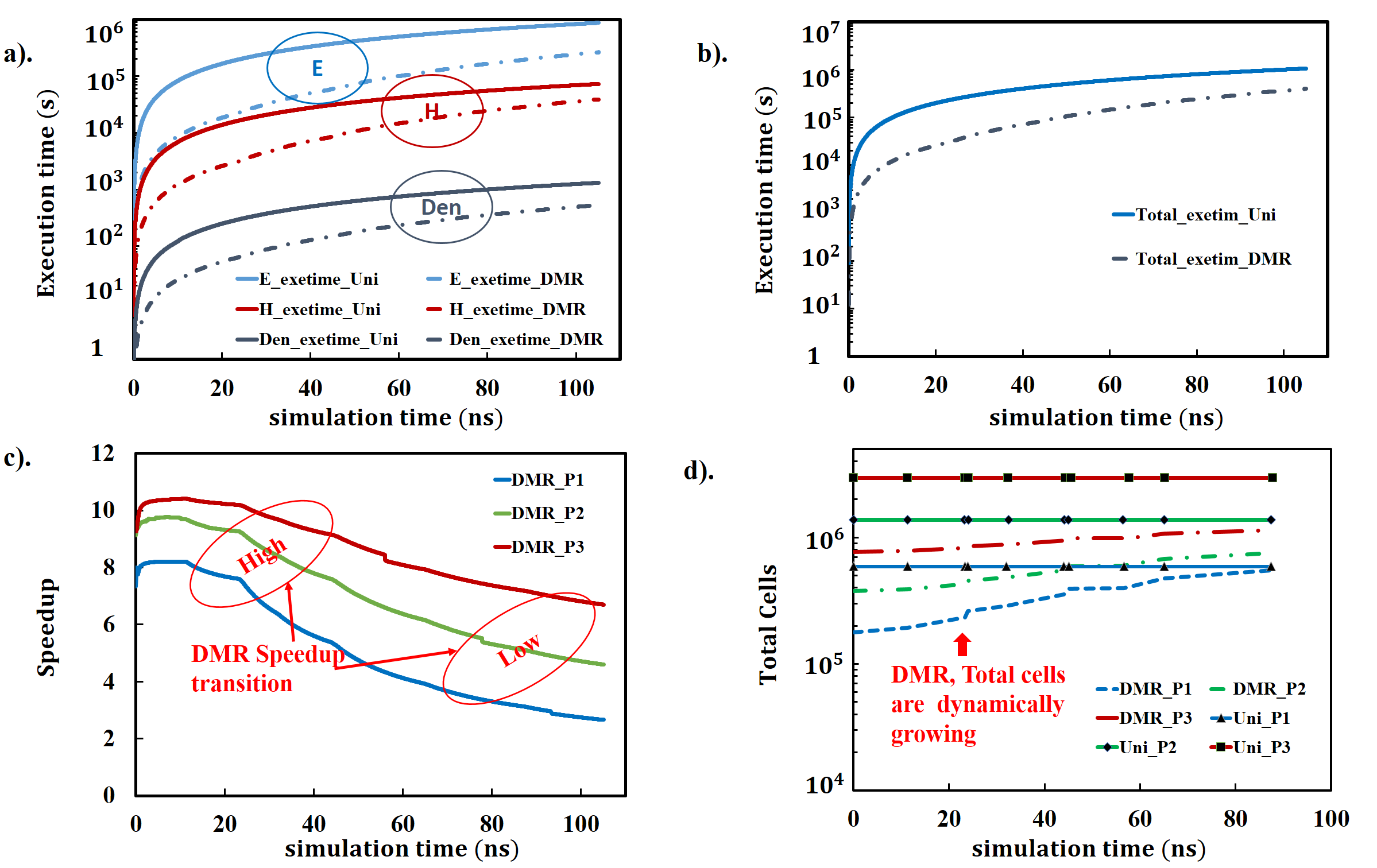}
\caption{For different simulation times in seconds using different techniques, Uni and DMR, the plots of (a). The subroutine wise Execution time to simulate P1, (b). The Total Execution time taken to simulate P1, (c). The Overall Speedup for different problem sizes, P1 to P3, and (d). The growth of total cells. Problem sizes, P1: $1.5\lambda\times1.5\lambda$, P2: $3.5\lambda\times1.5\lambda$, P3:$7.5\lambda\times1.5\lambda$. The subroutines E: E-field, H: H-field and Den: Plasma density. Uni:  uniform fine mesh and DMR: dynamic Mesh Refinement. The mesh refinement factor ($r$) of 2 is used.
}
\vspace{-2mm}
\label{subroutimtot}
\end{figure*}
For efficiency, subroutine-wise and overall execution time are compared between DMR and uniform fine mesh case (Uni) for a fixed problem size. Further, performance is evaluated for different problem sizes (P1, P2, and P3) to check the scalability of the proposed approach.
From Fig. \ref{subroutimtot} (a), it can be observed that the time required to execute the three constituent subroutines the E-field (E), H-field (H) and plasma density (Den) is higher for uniform fine mesh (Uni) in comparison to dynamic mesh refinement (DMR) for a fixed problem size (P1), P1: $1.5\lambda\times1.5\lambda$. 
The highest execution time is taken by E followed by H and Den. DMR significantly reduces the overall execution time as compared to Uni as observed in Fig. \ref{subroutimtot} (b).\\
\indent
Fig. \ref{subroutimtot} (a) and (b) shows that the DMR reduces the execution time with respect to the standard uniform mesh technique by a dynamic factor. This factor decreases as the simulation proceeds and the fine mesh expands. 
Fig. \ref{subroutimtot} (c) provides a better understanding on the contribution of DMR on the overall speedup and scalability for different problem sizes. For a fixed problem size, let P1: $1.5\lambda\times1.5\lambda$, the initial speedup is of the order of $8$ and drops to around $2$ as the simulation proceeds. The result is in agreement with Fig. \ref{subroutimtot}(b). For different problem sizes P2: $3.5\lambda\times1.5\lambda$ and P3: $7.5\lambda\times1.5\lambda$, the speedup transits from high to low and performs better as the problem size increases.   For different problem sizes P1 to P3, the overall speedup $\approx 5,\: 7\: \text{and}\: 8$, respectively. The higher performance for bigger problem sizes is due to very small value of $R(t)$ initially which grows gradually as plasma pattern spreads.
For uniform fine mesh (Uni) implementation, the total cells $\approx 10^{5}-10^{6}$  and total computations per iteration are always fixed (Fig. \ref{subroutimtot} (d)). Whereas, for DMR, total number of cells and computations grow as the refinement region grows with simulation time, $R(t)$. 
For uniform mesh, total computational cost is proportional to $N_{f}I_{f}$ (total cells x total iterations).
Whereas, for DMR, the total computational cost is proportional to sum of contributions from coarse and fine mesh. Coarse mesh has fixed number of cells ($N_{f}/r^2$) and updates are less frequent (by a factor of $r$) compared to fine mesh, therefore total cost for coarse mesh is $\big(1/r^{3}\big)N_{f}\:I_{f}$ (Total coarse cells $\times$ Coarse Iterations). For fine mesh, the computational cost is dynamic and is proportional to $R(t)\:N_{f}\:I_{f}$ (Total fine cells $\times$ Fine Iterations). Therefore, for DMR, we obtain that computational cost is proportional to $N_{f} I_{f} \bigg(1/r^{3}+R(t)\bigg)$, and primarily depends on refinement factor and refinement region. Higher the refinement factor and smaller the refinement region, better the speedup. 
Due to smaller number of cells in case of DMR (Fig. \ref{subroutimtot} (d)), size of data structure is initially small and the memory access performance is better compared to uniform fine mesh. Initially refinement region is very small and fine mesh data structure easily fits into cache memory which leads to higher cache hits and we observe a speedup better than the theoretical speedup of $r^3$ (Fig. \ref{subroutimtot} (c)).
\subsection{Physics of Complex plasma dynamics during HPM breakdown : spatio-temporal analysis}
\begin{figure*}[!htbp]
    \centering
    \includegraphics[width=0.7\textwidth, height=1.162\textwidth]{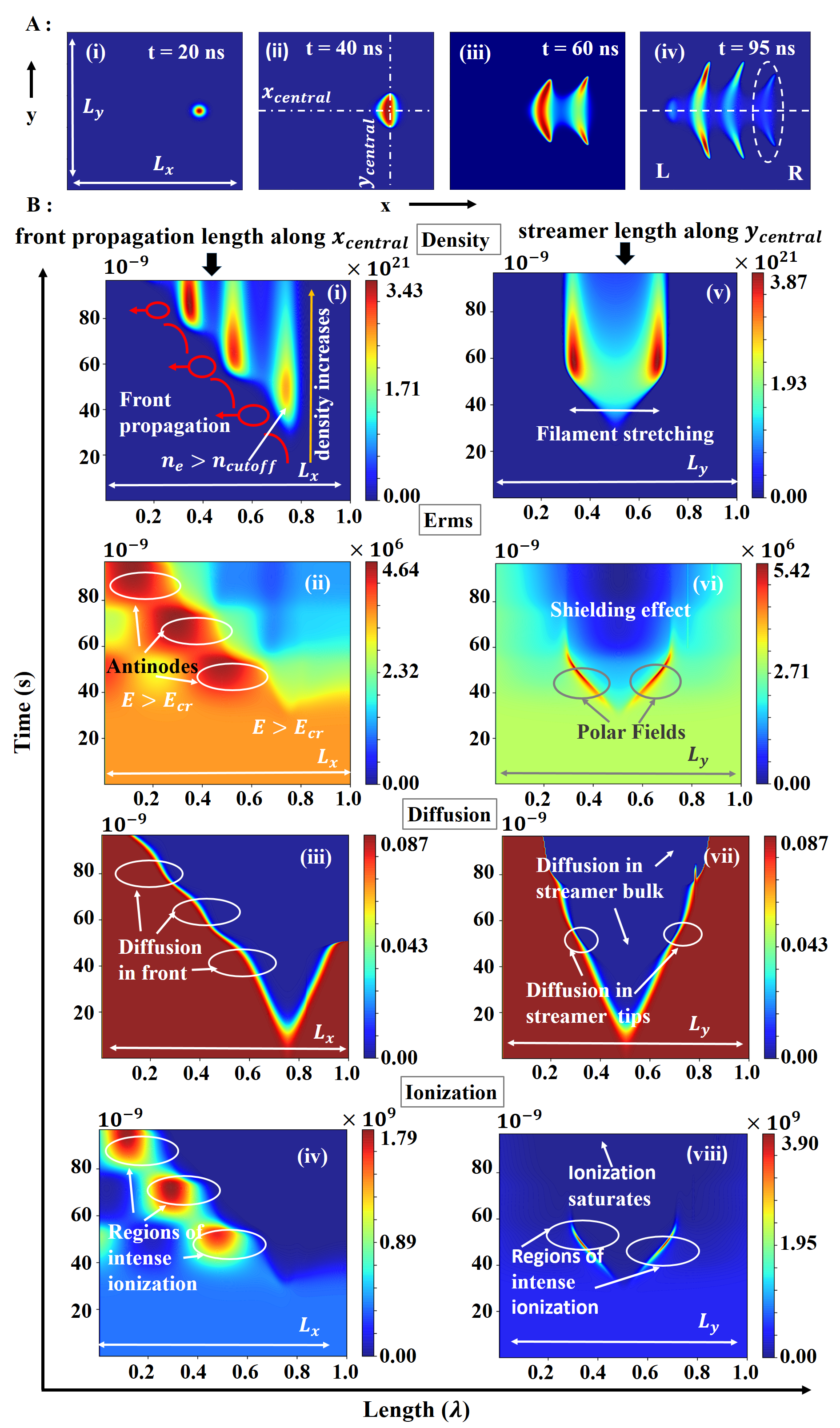}
    \caption{Part A: (i-iv) plasma density (m$^{-3}$) distribution in the filaments at four time instances using refinement factor ($r\:=\:2$).
    Part B: (i-iv) the evolution of the plasma density (m$^{-3}$), rms E-field (V/m), the effective Diffusion coefficient (m$^{2}$/s) and the rate of ionization (s$^{-1}$) along the $x_{central}$ of the computational domain. Part B: (v-viii) the evolution of the same quantities along the $y_{central}$ passing through the center of the rightmost filament (in dotted circle). Here, L: Left and R: Right, $L_{x}\:=\:L_{y}\:=\:1\lambda$.
    }
    \label{fig:plasmoidevol}
\end{figure*}

For the spatio-temporal analysis of plasma dynamics using DMR, we consider $\{ c_{kx},c_{ky}\}\:=\:\{1,1\}$ and $n_{0}$ is located at $0.75\:L_{x}$ and $0.5\:L_{y}$. $E_{0}$ is $5.0$ MV/m and wave frequency is 110 GHz.
When a powerful millimeter-wave with a strong E-field interacts with the air (or gas) at high pressure, it delivers sufficient energy to the free electrons that accelerate and causes continuous ionization of the air/gas and leads to an avalanche breakdown (when ionization overcomes the attachment, diffusion, and recombination). The process results in the plasmoid formation\cite{BChaudhury2010, Konstantinos-2015}. We can observe the plasmoid formation in Fig. \ref{fig:plasmoidevol} part A: (i), its growth into consecutive 1st, 2nd, and 3rd filaments and the propagation of the plasma filamentary structure from right to left with time in Fig. \ref{fig:plasmoidevol} part A: (ii),(iii) and (iv) respectively.\\
\indent
The simulation results in Fig. \ref{fig:plasmoidevol} part B: (i-iv) show the temporal evolution of plasma density, rms E-field, diffusion coefficient, and rate of ionization along the central x-axis of the computational domain, whereas Fig. \ref{fig:plasmoidevol} part B: (v-viii) represents the same quantities along the y direction (y-axis passing through the center of the rightmost filament). Fig.\ref{fig:plasmoidevol} helps us in capturing the complete spatio-temporal evolution of the plasmoid into filaments and its propagation during the first 100 ns which has never been reported in similar fashion and in so much details in earlier literature. \\
\indent
The initial plasmoid is located at $0.75\lambda$ from the leftmost boundary as shown in Fig. \ref{fig:plasmoidevol} part A: (i) which is captured at 20 ns. It remains Gaussian until the plasma density is small and cannot perturb the incoming microwave E-field as observed in Fig. \ref{fig:plasmoidevol} part B: (ii) \cite{BChaudhury2011}. The plasma density growth over time (Fig. \ref{fig:plasmoidevol} part B: (i)) can be attributed to the high incident rms E-field ($E_{\text{rms}}=3.5$ MV/m)$>E_{\text{cr}}$, where critical (or breakdown) field $E_{\text{cr}}\approx 2.5$ MV/m rms \cite{BChaudhury2010}. From Fig. \ref{fig:plasmoidevol} part A: (ii), we can observe how the E-field distribution along the central x-axis changes during the time duration 20 to 50 ns and during the same time interval, from Fig. \ref{fig:plasmoidevol} part B: (iv), we can observe how the rate of ionization increases from $0.18\times10^9$ s$^{-1}$ to $1.3\times10^{9} $ s$^{-1}$ along x-direction. From Fig. \ref{fig:plasmoidevol} part B: (ii) and (iv), we can clearly see the strong co-relation between the E-field peaks and the increased rate of ionization in space and time. This is because the high ambient pressure results in multiple collisions between the electrons and neutral gas species. The ionization being a collision assisted process gets governed by the local E-field or $E_{\text{eff}}$.
In our study at $f = 110$ GHz, the wave angular frequency ($\omega$) is $2\pi f\approx 6\times 10^{11} s^{-1}$, $\nu_m\approx4\times10^{12} s^{-1}$ , when $E_{\text{rms}}\geq3.5$ MV/m, then the $E_{\text{eff}}\approx3.46$ MV/m $\geq E_{\text{cr}}$, also satisfies the breakdown criteria. Thus, high collisions result in much more ionization which in turn increases the plasma density. The increase in the rate of ionization causes an increase in the density of the plasma bulk and the plasma starts modifying the incident field like a dielectric. It is also important to consider the growth of the filament along the vertical (y) axis Fig. \ref{fig:plasmoidevol} part B: (v-viii). From Fig. \ref{fig:plasmoidevol} part A: (i) and part B: (v) between t = 20 to 50 ns duration, we can see that the plasma density starts growing and the filament stretches along the vertical cross-section. The filament growth along the vertical direction can be attributed to the very large polar field that increases from $2.7$ MV/m to a maximum $\approx 5.42$ MV/m (Fig. \ref{fig:plasmoidevol} part B: (vi)), that results from Electrostatic polarization~\cite{BChaudhury2011}. 
The E-field in the center and at the tip of the filaments starts oscillating as the filament length approaches half wavelength. As discussed in \cite{BChaudhury2011} a resonance-like behavior may result in the streamer that starts acting as a half-wave dipole and reflects the incoming microwave E-field. We can observe the polar fields in Fig. \ref{fig:plasmoidevol} part B: (vi). The field enhancement results an intense ionization ($\approx 2$ times the ionization along the x-direction) as can be seen in Fig. \ref{fig:plasmoidevol} part B: (iv) and (viii). The high polar fields stretch the plasma filaments along the vertical direction (vertical polarization of incident microwave E-field ) as a result of higher ionization and the free electrons in the filament tips start diffusing with diffusion coefficient ($D_e$). The filament growth is related to the  plasma tip elongation rate given by, $v_{\text{streamer}}=2\sqrt{D_e \nu_i}$. The ionization frequency, $\nu_i$ has a strong dependence on the polar rms E-fields, therefore, the streamer tip velocity also enhances due to the high E-field. In our study, we obtain the streamer (or filament) elongation velocity, $v_{\text{streamer}}\approx 30$ km/s, which is $\approx 15$ km/s along each tip of the filament. The filaments achieve high streamer velocity, $\approx 30$ km/s, in the time duration 20 to 60 ns.\\
\indent
At t$>40$ ns, the plasma density, $n_e \approx 2 \times 10^{21}$ reaches well above the cutoff density. 
When plasma density crosses the cutoff density,
(here $n_{\text{cutoff}}\approx9\times10^{20}$ m$^{-3}$), plasma starts reflecting the EM wave \cite{BChaudhury2010}. We can observe from Fig. \ref{fig:plasmoidevol} part B:(i) and also in the part B: (v) at $t>40$ ns, the $n_{e}>n_{\text{cutoff}}$, thus the plasma filament/streamer transits from a dielectric behavior to a conductor. The scattered and the incident E-field of the EM wave interferes and form nodes and antinodes of the standing wave ahead of the first filament. The E-field has its node (minima) at the streamer center and the antinode (maxima) at $\approx 0.25\lambda$ from the initial plasmoid center ( $0.75\lambda$ from the left boundary), and it corresponds to the location where the new plasmoid starts to form. We can validate from the Fig. \ref{fig:plasmoidevol} part B:(ii) that at the antinodes the E-field strength is of the order of $\geq 4$ MV/m. The high field results in very intense ionization at the edge of the plasma filament. Therefore, the electrons at the filament edges starts to diffuse out with an effective diffusion coefficient, $D_{\text{eff}}\approx 0.087$ m/s$^{2}$ as can be seen in Fig. \ref{fig:plasmoidevol} part B: (ii-iv). It is interesting to note that the filament edge where the plasma density is low always diffuses very fast and forms a narrow diffusion channel called the plasma front. The plasma front propagates along the x-direction (right to left of computation domain) with a theoretical front velocity, $v_{\text{front}}=2\sqrt{D_{e} \nu_i}$, here, $v_{\text{front}}\approx 20$ km/s. We can observe that the streamer elongation is faster than the front propagation due to a higher E-field in the tip of plasma filaments as compared to the E-field at the antinode of the standing wave ahead of the filament. Both, the plasma front and the streamer diffuse with similar diffusion coefficients, $D_{\text{eff}}\approx D_e$, but the filament tip has higher density in comparison to the front and the growth of the streamer tip gets modulated with the polar fields of the streamer. \\
\indent
For t$>50$ ns, the second plasmoid follows a similar physics to elongate into a streamer. It starts perturbing the incoming E-field by obstructing the E-field to reach the initial (first) filament.
The high density in the second filament results in high plasma conductivity. Also, the width of the streamer almost nears the skin depth in plasma \cite{BChaudhury2010, BChaudhury2011}. Therefore, the shielding of the microwave occurs by the second filament on the first filament towards the right of the simulation domain. This we can observe in Fig. \ref{fig:plasmoidevol} part B: (v, vi and viii). The shielding effect obstructs the filament stretching in the vertical direction. Subsequently, the ionization along the filament tips gets saturated and stops the streamer elongation. The newly formed filament acts as the reflector of the incoming microwave.
The process of standing wave formation and enhancement of the front E-field follows and results in continuous ionization and diffusion-assisted front propagation. Further, it continues till the end of simulation (around 95 ns) and we observe a self-organized fish-bone like plasma filamentary structure exactly similar to experimental observations~\cite{Hidaka2008,Fukunari2019}.
\section{Conclusion}
In this paper, a dynamic mesh refinement (DMR) technique has been proposed to solve the computationally challenging Maxwell-plasma fluid model for simulating HPM breakdown induced plasma pattern formation and associated dynamics. The DMR technique leads to generation of two meshes, coarse mesh which is present throughout the computational domain and the fine mesh which evolves with time in a self-aware fashion depending on plasma and electric field gradients.
The implementation of the DMR technique has been described in details. Further the technique has been evaluated in terms of accuracy and speedup by applying it to simulate and reproduce the experimental observations under similar conditions. The technique could accurately reproduce the complex plasma dynamics and plasma structures for different problem sizes and refinement factors. For a refinement factor of 2, we obtain an overall speedup of $5$ to $8$ times for different case-studies. The bigger problem sizes, involving physical duration of $>100$ ns, which typically take months using an uniform fine mesh can be handled in few days using the proposed DMR technique. Finally, through a novel spatio-temporal visual analysis, the complex physics involved during the high power high frequency microwave breakdown and self-organized filamentation process have been discussed . The proposed DMR based technique will be very helpful to investigate longer time scale phenomena in HPM breakdown (in the order of micro to milliseconds) such as gas heating which is sparsely reported in the existing literature. DMR aided simulations will also help to efficiently investigate and better understand the different plasma structures observed in experiments.
\section*{Acknowledgment}
P. Ghosh would like to thank the DST, Gov. of India, for research fellowship received under DST-SERB project (Project No. - CRG/2018/003511).
\bibliographystyle{abbrv}
\bibliography{references}
\end{document}